\documentclass[aps,prl,preprint,superscriptaddress]{revtex4-1}
\usepackage{graphicx}
\usepackage[utf8]{inputenc}
\usepackage{textcomp}

\begin{document}

\title{Dissipation mechanism above the current threshold in Josephson
junction chains}

\author{Roland Schäfer}
\email{Roland.Schaefer@kit.edu}
\affiliation{Karlsruhe Institute of Technology, Institut für
Festkörperphysik, D-76021 Karlsruhe, Germany}
\author{Wanyin Cui}
\affiliation{Karlsruhe Institute of Technology, Institut für
Festkörperphysik, D-76021 Karlsruhe, Germany}
\affiliation{Karlsruhe Institut of Technology, Physikalisches Institut,
D-76131 Karlsruhe, Germany}
\author{Kai Grube}
\affiliation{Karlsruhe Institute of Technology, Institut für
Festkörperphysik, D-76021 Karlsruhe, Germany}
\author{Hannes Rotzinger}
\affiliation{Karlsruhe Institut of Technology, Physikalisches Institut,
D-76131 Karlsruhe, Germany}
\author{Alexey V. Ustinov}
\affiliation{Karlsruhe Institut of Technology, Physikalisches Institut,
D-76131 Karlsruhe, Germany}


\date{October 15, 2013}

\begin{abstract}
We present measurements on one-dimensional Josephson junction arrays formed
by a chain of SQUID loops in the regime where $E_J / E_c\simeq 1$.  We
observe a blockaded zero current branch for small bias voltages. Above a
certain voltage $V_\mathrm{sw}$ the $I$-$V$ characteristics changes
discontinuously to a dissipative branch characterized by a flux dependent
conductance. Three of four samples show a pronounced hysteresis for
a forward and backward voltage sweep. We observe a periodic field
dependence of the conductance above $V_\mathrm{sw}$ which can be described
with $P(E)$ theory for cooper-pairs and therefore gives evidence of viscous
dynamics of Cooper pair transport through the array.
\end{abstract}

\pacs{}

\maketitle

The physics of small single superconducting islands connected via the Josephson
effect to an environment is well understood in various regimes and can be used
as a building block for more complex systems. For low dimensional arrangements
of these islands one finds remarkable similarities to other research areas.
For example, the suppression of the electrical conductance in one or two
dimensional ultra thin superconducting granular films \cite{goldman98} is also
observed in long arrays of superconducting nano-islands.  Another intriguing
example is the projected quantum mechanical duality of solitonic single fluxon
excitations observable in long Josephson contacts to the charge transport in
series of small capacitance superconducting islands
\cite{haviland96,hermon96,rachel09,homfeld11}.

\begin{figure}[h!]
(a)\begin{minipage}[t]{80mm}
\vspace{0pt}\par\includegraphics[origin=t]{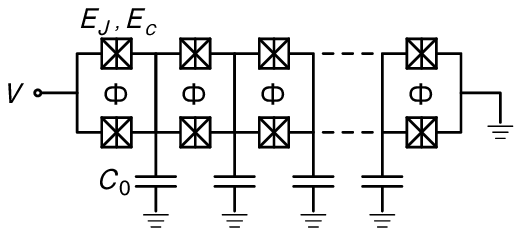}
\end{minipage}
(b)\begin{minipage}[t]{80mm}
\vspace{0pt}\par\includegraphics[origin=t]{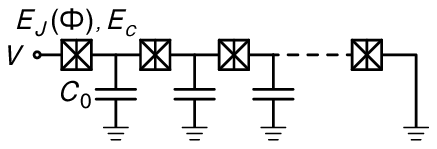}
\end{minipage}
(c)\begin{minipage}[t]{80mm}
\vspace{0pt}\par\includegraphics[width=80mm,origin=t]{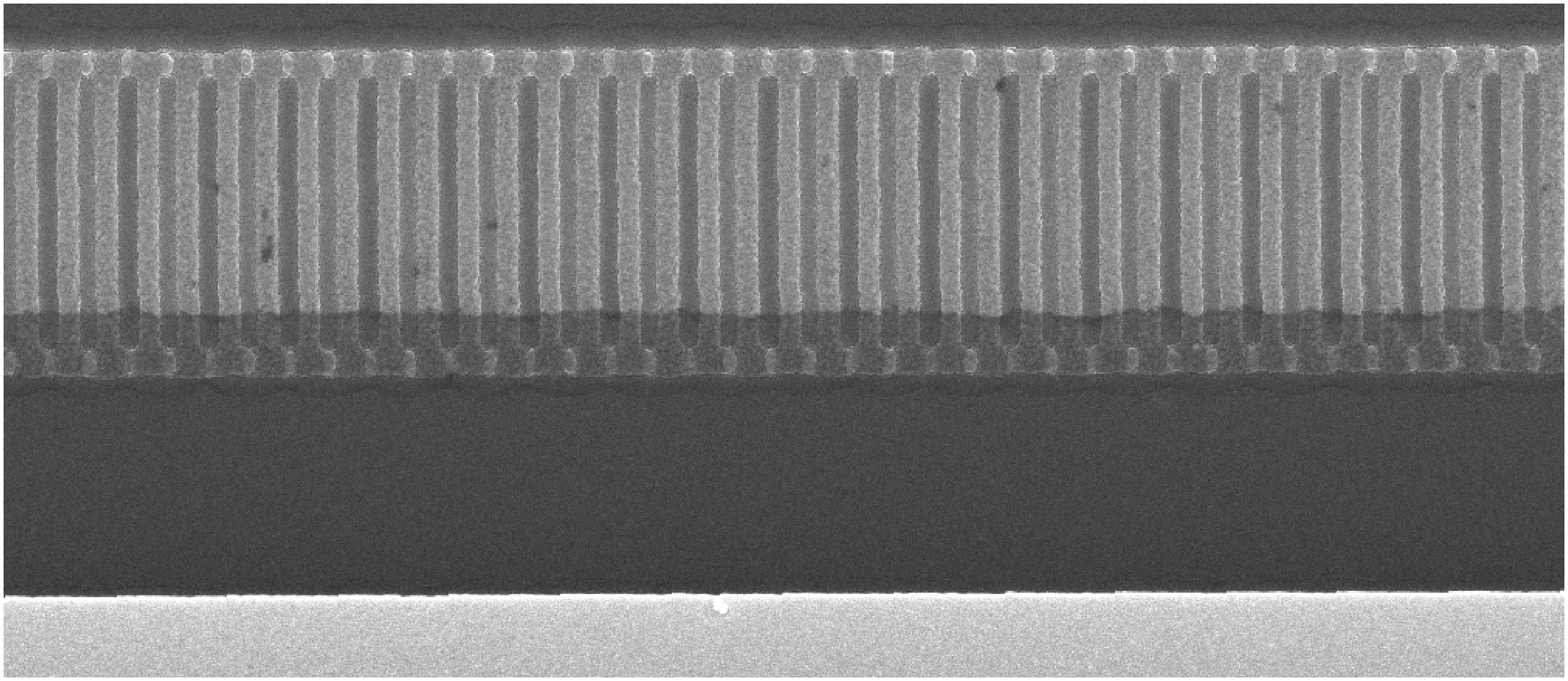}
\end{minipage}
\caption{\label{fig:i}Schematic representation of investigated samples. (a)
Each Josephson junction is characterized by the coupling energy $E_J$ related
to the Josephson coupling and the charging energy $E_c$ related to the junction
capacity. The chains are formed by two parallel strands linked between
junctions. SQUID loops form which are penetrated by a magnetic flux $\Phi$. The
links define islands which have in addition to the junction capacitances
(quantified by $E_c$) a capacitance $C_0$ to ground. (b) Simplified view: The
SQUID loops in (a) can be modeled as single junctions with an effective
coupling $E_J(\Phi)$ which periodically modulates with the flux (see Eq.
\ref{eq:ejeff}). (c) scanning electron micrograph showing a detail of the JJ
array. The bone shaped aluminum structures overlap at the top and bottom and
form the Josephson contacts in this region. The complete JJ array comprises 255
SQUID loops in total. The pitch between adjacent JJ measures 200\,nm. The
extend of the array in vertical direction is $1.6\,$\textmu{}m. At the bottom
of the micrograph the capacitively coupled ground plane dominating $C_0$ can be
spotted.}
\end{figure}

In this paper we report on measurements on nano-structured one dimensional
chains of superconducting islands, connected via Josephson tunnel junctions
(JJ)  (for a sketch, see Fig.~\ref{fig:i}).  The islands are characterized by a
small self capacitance $C^\prime$ which is dominated by the capacitance of the
Josephson junctions and a small contribution $C_0$ ($C^\prime/C_0\sim10^{-2}$)
originating from the metallic environment held at ground potential.  Due to the
size of $C^\prime\sim1$\,fF the corresponding charging energy of one island
$E_c=2e^2/C^\prime$ is large, $E_c\gg{}k_\mathrm{B}T$ at operation temperatures
below  0.1K.  Thermally activated processes are strongly suppressed. From a
conductance point of view, $E_c$ is in turn in competition with the Josephson
coupling energy $E_J$. In the limit $E_J/E_c\gg1$
the array behaves entirely superconducting, manifested, e.g., in a supercurrent
at zero voltage \cite{chow98,agren01}, whereas for
$E_J/E_c\le1$ the desired zero current state is observable at
finite voltages \cite{haviland96}.  From the normal state conductance above
$T_c$ the Ambegaokar-Baratoff relation\cite{ambegaokar63, *ambegaokar63e}
$E_J=g\Delta/(2N)$ is helpful for an estimate of the Josephson
coupling energy of on island to the next. Here $\Delta$ is the superconducting
gap energy, $g=G/G_q$ the conductance in units of $G_q=4e^2/h$ and $N$ is the
number of JJ. Both the capacitance and $g$ are proportional to the junction
area of a Josephson contact. 

For shadow-evaporated Al/AlO$_x$/Al tunnel junctions, employed in this work,
the regime $\Delta>E_c>E_J$ can easily be reached. Furthermore, it is
convenient to implement the Josephson junctions in the shape of SQUID loops as
indicated in Fig.\ \ref{fig:i}\,(a). Each SQUID loop is equivalent to a single
Josephson junction, but with $E_J$ tunable by an magnetic field
$B_\mathrm{ext}$. In the presented case, where the geometrical inductance $L$
of the loops can be neglected ($LI_c< 10^{-3}\Phi_0$, where $I_c$ is the
critical current of the junctions and $\Phi_0=h/e$), the effective Josephson
coupling energy reads (see, e.\,g., Ref.\ \onlinecite{tinkham})
\begin{equation}\label{eq:ejeff}
E_J(\Phi)=\sqrt{ (E_s^2-\delta^2)\cos^2(\pi\Phi/\Phi_0)+\delta^2 },
\end{equation}
where $E_s=E_1+E_2$, $\delta=|E_1-E_2|$, and $E_1,E_2$ are the coupling
energies of the individual SQUID junctions. For a perfectly symmetric loop size
$\delta=0$ and full suppression of the coupling can be expected at half-integer
flux quanta, $\Phi=(n+1/2)\Phi_0$. Furthermore, in a homogenous chain of SQUIDs
with identical loop area $A$ supression of the Josephson coupling would happen
at the same field values $B_\mathrm{ext}=(n+1/2)\Phi_0/A$ for all links.  In
the experimental realization of nano-scale Josephson tunnel junctions, the
unavoidable finite parameter spread in the loop area and the individual
junctions leads to a spread in $E_J(B_\mathrm{ext})$, so a complete
suppression and perfect periodicity of $E_J$ with the magnetic field cannot be
expected.

%
For this study four nominally identical JJ arrays have been fabricated on
the same SiO$_2$  isolated silicon substrate.  A scanning electron
micrograph of such an array is shown in Fig.~\ref{fig:i}\,(c). 

The arrays are fabricated using conventional electron beam lithography and
shadow-evaporation techniques \cite{dolan77,niemeyer74}. The shadow pattern
is similar to the one described in Ref.\ \onlinecite{haviland96}, with the
modification that the distance of the SQUID junctions has been extended to
about $1.3$\,\textmu{}m for larger flux sensitivity. Between the
shadow-evaporation steps an oxide barrier is grown in a pure oxygen
atmosphere of 3\,Pa for 5\,min. The array is connected to a pre-defined
wiring layer of an  Au-Pd alloy, not shown in the micrograph. 
The superconducting part of the structure is kept small to reduce influence
of non-equilibrium quasiparticles which harm many experiments which
rely on the freezing out of quasiparticle excitations at very low
temperatures.  For a well defined capacitive environment $C_0$ of each
island, we put two ground leads at a distance of $1.2$\,\textmu{}m on each
side of the array.

The junction capacitance $C^\prime$ and the coupling capacitance $C_0$ to
ground  are determined from the geometrical dimensions deduced from
scanning electron micrographs.  Assuming the value of 45\,fF/\textmu{}m$^2$
for the typical specific capacitance of Al/AlO$_x$/Al tunnel
junction\cite{haviland96} we estimate the mean self capacitance of the
islands to be $C=1.15\,$fF$\pm20$\%, where the error margin reflects both
systematic deviations as well as island-to-island variations. The
capacitance to ground is bounded by $5\,\text{aF}<C_0<20\,$aF, by
considering the minimal and maximal capacitance of a strip line of similar
geometry formed by our ground gate and a center conductor replacing the JJ
chain.  The screening parameter $\Lambda=\sqrt{C/C_0}$ specifying the
number of islands over which the electrostatic potential created by an
extra charge drops off is thus bound to $6<\Lambda<17.$

\begin{figure}[t!]
(a)\hfill(b)\hspace{40mm}\rule{0mm}{1ex}\\[2mm]
\includegraphics[height=36mm]{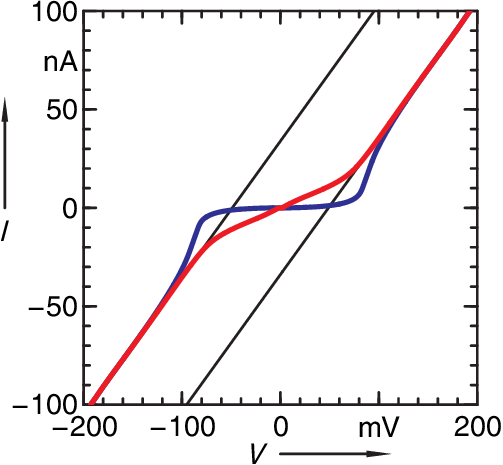}
\hfill\includegraphics[height=40mm]{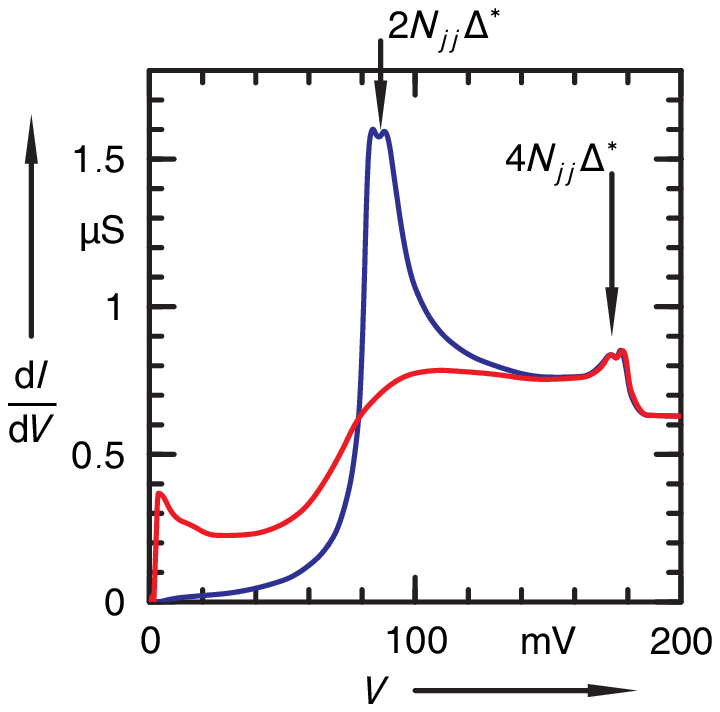}
\caption{\label{fig:iii}  (a) large scale $I$-$V$ characteristics of sample III
at zero field (red curve) and at half flux quantum penetrating each SQUID loop
(blue curve). (b) differential conductance of sample IV (same color coding as
in (a)). inset: data for sample III.}
\end{figure}

The normal-state resistance $R_h=1/g=N \overline{R}_{JJ}$ of the chain can
be derived from the linear slope of the $I$-$V$ characteristics at high bias
voltage (see Fig.~\ref{fig:iii}\,(a)). $\overline{R}_{JJ}$ is the mean
resistance across a single SQUID loop. The parameters for the four samples
are summarized in Tab.~\ref{tab:param}. Three of them display very similar
parameters while sample I has a much higher resistance.  In this paper we
focus on measurements for sample III. The results for sample II and IV are
very similar. At the end of the paper we briefly point out differences of
sample I to  sample II--IV. 
\begin{table}[b]
\caption{\label{tab:param} Parameter of the 4 samples.  $V_h$: Voltage
offset of the assymtotic behavior at large bias.  $R_h$: Differential
resistance at large bias.  $E_c=eV_h/4N.$ $E_J=\Delta h /(8
e^2R_h/N)$ is evaluated using the Ambegokar-Baratoff relation
assuming $\Delta=200\,$\textmu{}eV.  } \begin{tabular}{llllll}\hline
Sample & $V_h$ & $R_h$     & $E_J$       & $E_c$       & $E_J/E_c$ \\
       &  mV   & M\textohm & \textmu{}eV & \textmu{}eV &           \\
\hline
I      & 110.0 & 3.13      &  53         & 107         & 0.49      \\
II     &  97.3 & 1.32      &  125        & 95          & 1.31      \\
III    &  94.5 & 1.46      &  113        & 93          & 1.21      \\
IV     &  93.4 & 1.32      &  125        & 92          & 1.37      \\
\hline\end{tabular}
\end{table}

%
\begin{figure}[t!]
\includegraphics[width=86mm]{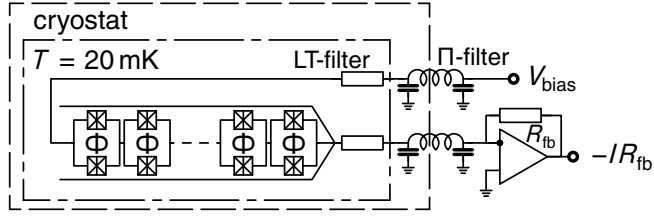}
\caption{\label{fig:ii} Schematics of the experimental setup.}
\end{figure}
The setup is summarized in Fig.~\ref{fig:ii}. The sample is mounted inside a
RF-tight copper box held at the base temperature ($T\simeq 10$\,mK) of a
commercial dilution refrigerator. All electrical connections of this box are
carefully filtered by metal powder filters (cutoff frequency $\sim100\,$MHz).
Together with lumped element $RC$ filters thermally anchored at the mixing
chamber they form low-temperature (LT) filters with a bandwidth of about
10\,kHz. All wires leading into the cryostat, which is by itself a metallic
cage, are equipped with $\Pi$-filters to keep the interior of the cryostat
RF-free.  For the $I$-$V$ characteristics the current is recorded with a
homemade transimpedance amplifier. The output voltage of the amplifier
$V_\mathrm{out}=-I_\mathrm{jj}R_\mathrm{fb}$ is measured with a digital
voltmeter, with a gate time of 200\,ms.  The voltage drop over the line
impedance $R_\ell$ is accounted for numerically to yield the actual bias on the
sample $V_\mathrm{bias}=V_\mathrm{in}-IR_\ell$.

%
\begin{figure}[t!]
\includegraphics[height=40mm]{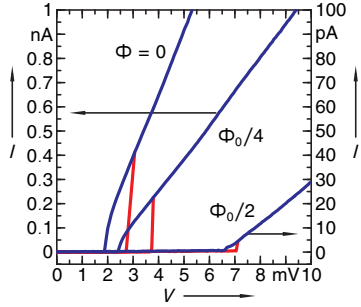}
\caption{\label{fig:iv} small scale $I$-$V$ characteristics of sample III
for three different values of the flux penetrating each SQUID loop.  The
characteristics display a hysteretic behavior at the onset of conductance. The
red curves are measured while the bias voltage is increased, the blue curves on
the way back.  Note that the characteristic for $\Phi=\Phi_0/2$ is scaled up by
a factor of 10 (right scale) as compared to the two other ones (left scale).}
\end{figure}

Fig.~\ref{fig:iii}\,(a) shows $I$-$V$ characteristics on a large bias voltage
scale up to $\pm200\,$mV measured at two different frustrations
$f\equiv\Phi/\Phi_0=0$ (red) and $f=1/2$ (blue). Intermediate magnetic fields
interpolate periodically between these curves.  The dependence on magnetic
field is largest below a voltage threshold which corresponds approximately to
$2N\Delta/e\sim100\,$mV. The steep rise at zero magnetic field, evident from
the blue curve in Fig.\ \ref{fig:iii}\,(a), can tentatively be identified with
the onset of pair breaking processes and the transport above $2N\Delta^\ast/e$
is dominated by quasiparticles. Here, the energy gap $\Delta^\ast$ can be
smaller than the zero temperature gap $\Delta$ of aluminum due to
non-equilibrium effects\cite{krasnov06}.

At lower bias voltages, the charge transport has to rely on the Josephson
coupling which is strongly suppressed at $f=n+1/2$. The periodic modulation of
the transport characteristics below $|eV|<2N\Delta$ is evidence that the main
transport mechanism in this voltage bias regime is due to the motion of Cooper
pairs.  However, the motion of Cooper pairs is presumably incoherent as it is
accompanied by dissipation.  The magnetic field dependence is much weaker at
$|V|>2N\Delta$, but still significant. Fig.\ \ref{fig:iii}\,(b) displays the
differential conductance for sample IV. Here the field dependence above
$2N\Delta$ is clearly visible. Only above an energy scale set approximately by
$4N\Delta$ the field dependence is below the experimental resolution.  Sample
II and III (data not shown) show a very similar crossover from magnetic
field-dependent to magnetic field-independent.

At voltages $100\,\text{mV}<|V|<180\,$mV the $I$-$V$ characteristics approach
approximately straight lines shown in black in Fig.\ \ref{fig:iii}\,(a). The
lines for positive and negative voltages display a horizontal distance of
$V_h=98$\,mV. This value is in approximate agreement with
$4NE_c/e=75\,$\textmu{}V where $E_c$ has been estimated from the geometry of
the contacts. $4NE_c$ is the most simple estimate of the asymptotic offset
voltage due to the Coulomb blockade of single charge transport. According to
Fig.\ \ref{fig:iii}\,(b) the asymptotes at high voltage bias have a slightly
lower slope than the estimates based on the intermediate voltage range.  The
features in Fig.\ \ref{fig:iii}\,(b) at $2N\Delta^\ast$ and $4N\Delta^\ast$
have remarkable similarities with stacks of intrinsic Josephson junctions in
HTC-cuprates\cite{krasnov06}. These have been explained by nonlinear
non-equilibrium quasiparticle relaxation, which is also reasonable to assume
here. The lower slope above $4N\Delta^\ast$ can be interpreted as sign of a
lower quasiparticle density due to recombination.

For the remaining part of this paper we focus on the properties of $I$-$V$
characteristics at much smaller voltages. Fig.\ \ref{fig:iv} gives three
examples at different values of the magnetic flux penetrating each SQUID
loop. In each case a hysteresis in the $I$-$V$ characteristic is found. For
other values of $\Phi$ the behavior interpolates periodically between the
shown examples. At zero voltage bias the differential resistance approaches
a value above the measurement limits set mainly by the current resolution
$I_\mathrm{min} \sim 50\,\mathrm{fA}$ (rms) of the setup to $R_\mathrm{max}
\sim 100\,$G$\Omega$. On increasing the bias the blockade of Cooper pair
transport prevails up to a switching voltage $V_\mathrm{sw}$, see sharp
increase of the conductance colored red in Fig.~\ref{fig:iv}. For higher
bias the $I$-$V$ dependence follows a (dissipative) straight branch which
extends on lowering the bias to voltages below $V_\mathrm{sw}$. In the
backsweep of the bias voltage the chain retraps in a continuous manner at a
well defined $V_\mathrm{rt}$; no discontinuous "jumps" could be detected.  

\begin{figure}[t!]
\includegraphics[height=49mm]{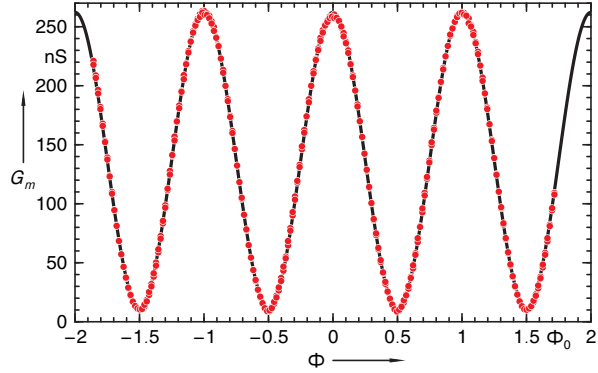}
\caption{\label{fig:v} Differential resistance above switching as function of
magnetic flux (dots). The solid line is given by
$g(\Phi)=(G_m(0)-G_m(\Phi_0/2))cos^2(\pi\Phi/\Phi_0)+G_m(\Phi_0/2)$ (see
text).}
\end{figure}
Figure \ref{fig:v} shows the measured slope $G_m$ of the $I$-$V$ curves above
the switching voltage as a function of the applied flux.  In this range the
differential conductance is in good approximation independent of the bias
voltage (see Fig.~\ref{fig:iv}) and thus gives a measure for the strength of
dissipation.  In Fig.\ \ref{fig:v} we underlay a function of the form
$(G_m(0)-G_m(\Phi_0/2))\cos^2(\pi\Phi/\Phi_0)+G_m(\Phi_0/2)$ (black line,
$G_m(0)\simeq259\,$nS and $G_m(\Phi_0/2)\sim9.5\,$nS) to the measured data
points. 

In comparison with Eq.\ (\ref{eq:ejeff}) it is evident that $G_m(\Phi)$ is
proportional to $E_J^2(\Phi)$.  This is our main result. It is rather
surprising that the effective Josephson coupling of a single superconducting
loop gives already a good description of the dissipative conductance branch of
a chain of loops. However, very little is known on the collective effects of
the simultaneous transport of cooper pairs and quasiparticle excitations along
a chain of small Josephson junctions.

\begin{figure}
\includegraphics[height=49mm]{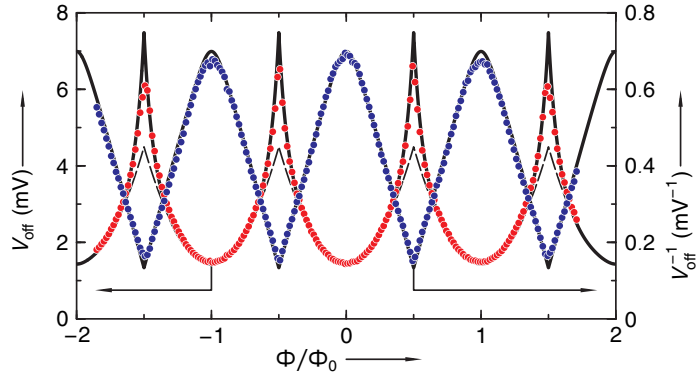}
\caption{\label{fig:vi} The offset voltage $V_\mathrm{off}$ (Red dots)  as
function of frustration $f=\Phi/\Phi_0$.  Data have been extracted from $I$-$V$
characteristics. The lines are explained in the text. Blue dots: The inverse of
$V_\mathrm{off}$ (right scale).}
\end{figure}
The linear behavior above switching evident from Fig.\ \ref{fig:iv} defines a
further voltage $V_\mathrm{off}$ at which these straight lines cross the
abscissa of zero current.  In Fig.\ \ref{fig:vi} we present empirical findings
for $V_\mathrm{off}$ as a function of the frustration $f$.  The measured data
(red dots) are well approximated by 
\begin{equation}\label{eq:hyperoff}
V_\mathrm{off}^{-1}(f)= -A\sqrt{B^2+f^2}+C.
\end{equation}
A least square fit yields $A=1.7$\,mV, $B=0.07$, and $C=0.79$\,mV. This
functional dependence is shown as solid line in Fig.\ \ref{fig:iv}.  Equation
(\ref{eq:hyperoff}) expresses the following experimental findings:
$1/V_\mathrm{off}$ plotted as a function of $f$ (blue dots in Fig.\
\ref{fig:iv}) has a hyperbolic shape; it follows a quadratic behaviour in the
vicinity of $f=0$ (more generally close to $f=n$). On increasing $f$ it quickly
approaches a straight line with a slope of $A=\pm{}1.7\,$mV. At maximal
frustration ($f=(n+1/2)$), we find a finite value of about
$V_\mathrm{off}\sim7$\,mV and the slope in $1/V_\mathrm{off}$ switches sign.

Figures \ref{fig:v} and \ref{fig:vi} display explicit results for sample III.
The other, nominally identical samples on the same chip show a very similar
behaviour with differences only in small details.  For samples II and IV we
observe a splitting of the minima in $G_m(\Phi)$ around $\Phi=\pm1.5$. Similar
effects have been reported by other authors (see, e.\,g., Ref.\
\onlinecite{haviland96}) and can be attributed to typical artefacts of the
fabrication method.  Sample I displays a similar blockade of current at small
bias and a sinusoidal modulation of $G_m(\Phi)$, too, although $G_m(0)$ is by a
factor of 2 smaller as compared to the other samples.  However, it lacks the
hysteresis observed for the other samples and displayed in Fig.\ \ref{fig:iv}
for sample III. 

%
A reliable theory for the system investigated in this paper is still missing.
The system is difficult to describe theoretically because the interplay between
Cooper pair and quasiparticle tunneling on the one hand---different rules apply
in both cases---and the conversion between these distinct species of charge
carriers on the other hand.  Here, we present a simple phenomenological
picture, which is in good agreement with our findings. 

The chain of superconducting islands as a whole is voltage biased. The shunt
build by the stray capacitance of the leads in parallel to the chip capacitance
of the low temperature LP-filters (see Fig.\ \ref{fig:ii}) stabilizes the
voltage drop across the chain; in case of tunneling, the voltage drop does not
change significantly. On the contrary, a current biased setup would require a
Thévenin's equivalent impedance large compared to
$R_q=h/(4e^2)\sim6\,\text{k}\Omega$ at the relevant frequencies, which is as
high as the plasma frequency of the Josephson junctions $\omega_p\sim450\,$GHz.
In practise, this requirement is seldom fulfilled because any stray capacitance
in the vicinity of he array will result in a low impedance environment at high
frequencies.

For junctions deep inside of the array the situation appears different.  Its
environment can be viewed as a transmission line made from a chain of lumped
element units. Each unit contains a Josephson element and $C_0$ and its
transmission properties are dominated by the Josephson inductance
$L=\Phi_0/(2\pi{}I_c)$ shorted by $C_0$ to ground. A chain of such units has a
typical impedance of $Z(\omega)\sim\sqrt{L/C_0}\sim20R_q$ for all frequencies
below $\omega_p$. This argument shows that each junction of the chain is
effectively current biased although the device as a whole is connected to a
voltage source. Furthermore, to a good approximation individual junctions are
embedded into an environment of predominately ohmic impedance at all relevant
frequencies. 

The situation of a Josephson junction in a resistive environment is well
described by $P(E)$-theory \cite{leshouche2}  which yields for the tunneling
rate of Cooper pairs:
\begin{equation}\label{eq:pofe}
\Gamma=\frac{\pi}{\hbar}E_J^2P(2eV),
\end{equation}
\[P(E)=\frac{1}{h}\int_{-\infty}^{\infty}\mathrm{d}t\, e^{J(t)}e^{iEt/\hbar}\]
and
\[J(t)=2\int_{-\infty}^{\infty}\frac{\mathrm{d}\omega}{\omega}
\frac{\Re(Z(\omega))}{R_Q}\frac{e^{-i\omega t}-1}{1-e^{-\beta\hbar\omega}}.\]
$P(E)$ is the probability of the appropriate energy exchange between the
tunneling Cooper pair and the electromagnetic environment of the Josephson
junction. $J(t)$ is an equilibrium phase-phase correlation function which can
be expressed via the fluctuation-dissipation theorem as an integral over the
real part of the complex impedance $Z(\omega)$, where $Z(\omega)$ is the
impedance of the environment of the Josephson junction.  The essence of Eq.\
(\ref{eq:pofe}) is the proportionality between the incoherent tunneling rate
and $E_J^2$. It is a direct consequence of Fermi's golden rule used in the
theory to derive Eq.\ (\ref{eq:pofe}). In the present case of
$\Re(Z(\omega))>R_q$, P(E) is expected to peak at $E_c$: For junctions inside
the array tunneling of Cooper pairs has a high probability at a voltage drop of
order of $E_c/2e$. The voltage on each junction is increasing with a rate
$\dot{V}=I/C$ until a tunneling event at around $V_c=E_c/2e$ leads to a sudden
jump by $\Delta V=-E_c/e$.  For different junctions in the chain the voltage
will be different. The mean voltage will adjust itself in a dynamical manner
depending on local details and is increasing proportional to the current due to
the incoherent nature of the assumptive process.

%
In conclusion we have demonstrated that the current in an one-dimensional array
of small capacitance Josephson junctions just above the threshold voltage is
mainly carried by incoherent tunneling of cooper-pairs and follows a
characteristic $E_J^2$ behavior. This is supported by the $P(E)$-theory, in the
situation where one dominating superconducting island is embedded in an high
impedance environment of many islands.

\bibliography{rschetal}

\end{document}